\documentclass[conference]{IEEEtran}
\IEEEoverridecommandlockouts

\usepackage{diagbox}
\usepackage{booktabs}
\usepackage{colortbl}
\usepackage{multirow}
\usepackage{tabularx}
\usepackage{soul}
\usepackage{wrapfig}
\usepackage[caption=false]{subfig}
\usepackage[font=footnotesize,labelfont=bf]{caption}
\usepackage{dblfloatfix}
\usepackage{enumitem}
\usepackage{tablefootnote}
\usepackage{makecell}
\usepackage{pifont}
\usepackage[colorlinks=true, urlcolor=black, linkcolor=black, citecolor=black, pdfborder={0 0 0}]{hyperref}

\newcommand{\cmark}{\ding{51}}
\newcommand{\xmark}{\ding{55}}

\usepackage{flushend}

\usepackage{cite}
\usepackage{amsmath,amssymb,amsfonts}
\usepackage{algorithmic}
\usepackage{graphicx}
\usepackage{textcomp}
\usepackage{xcolor}
\def\BibTeX{{\rm B\kern-.05em{\sc i\kern-.025em b}\kern-.08em
    T\kern-.1667em\lower.7ex\hbox{E}\kern-.125emX}}
    
\begin{document}

\title{{\huge{\texttt{RTL++}: Graph-enhanced LLM for RTL Code Generation}}}

\author{\IEEEauthorblockN{Mohammad Akyash, Kimia Azar, Hadi Kamali}
\IEEEauthorblockA{\textit{Department of Electrical and Computer Engineering (ECE), University of Central Florida, Orlando, FL 32816, USA} \\
\{mohammad.akyash, azar, kamali\}@ucf.edu}
}

\maketitle

\begin{abstract}
As hardware design complexity escalates, there is an urgent need for advanced automation in electronic design automation (EDA). Traditional register transfer level (RTL) design methods are manual, time-consuming, and prone to errors. While commercial (instruction-tuned) large language models (LLMs) shows promising performance for automation, they pose security and privacy concerns. Open-source models offer alternatives; however, they frequently fall short in quality/correctness, largely due to limited, high-quality RTL code data essential for effective training and generalization. This paper proposes \texttt{RTL++}, a first-of-its-kind LLM-assisted method for RTL code generation that utilizes graph representations of code structures to enhance the quality of generated code. By encoding RTL code into a textualized control flowgraphs (CFG) and data flow graphs (DFG), \texttt{RTL++} captures the inherent hierarchy, dependencies, and relationships within the code. This structured graph-based approach enhances the context available to LLMs, enabling them to better understand and generate instructions. By focusing on data generation through graph representations, \texttt{RTL++} addresses the limitations of previous approaches that rely solely on code and suffer from lack of diversity. Experimental results demonstrate that \texttt{RTL++} outperforms state-of-the-art models fine-tuned for RTL generation, as evaluated using the VerilogEval benchmark’s $Pass@1/5/10$ metric, as well as the RTLLM1.1 model, which highlight the effectiveness of graph-enhanced context in advancing the capabilities of LLM-assisted RTL code generation\footnote{Dataset/Model is available at \NoHyper\cite{rtlpp_hugginface}.}.

\end{abstract}


\section{Introduction}

Large language models (LLMs) like GPT have shown exceptional capabilities in natural language processing (NLP) \cite{brown2020languages}, driving interest in their applications beyond traditional NLP tasks, particularly code generation \cite{nijkamp2023code}. Commercial LLMs like Codex \cite{chen2021evaluating}, AlphaCode \cite{li2022competition}, PaLM2 \cite{anil2023palm2technicalreport}, and Claude \cite{claude2023}, have significantly advanced software development by understanding and generating code\footnote{It is due to the abundance of high-quality training data, well-defined patterns, and mature ecosystems in languages like C++/Python \cite{cui2024origen}.}. While generating RTL code from natural language representation of design functionality/architecture descriptions can boost efficiency in hardware development (by reducing the effort for manually RTL coding, testing, and verification), their effectiveness in hardware design, particularly for register transfer level (RTL) remains constrained due to several challenges: (i) the lack of reliable training dataset, (ii) limited understanding of concurrent nature of RTL by LLMs, (iii) no consideration of resource constraining, etc. \cite{akyash2025simeval}. 

\begin{figure}[b]
\centering
\vspace{-15pt}
\includegraphics[width=\linewidth]{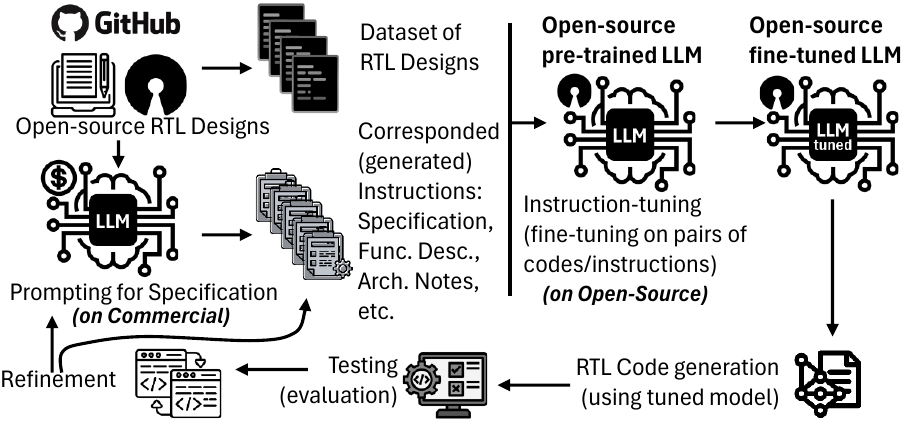}
\caption{The overview of Fine-tuning and (Potential Directions of) Refinement using Open-Source LLMs for Specializing of RTL Code Generation.}
\label{fig:llm_rtl_tuning}
\end{figure}

Recent advancements in commercial LLMs have demonstrated substantial improvements in RTL generation \cite{liu2024rtlcodera}. However, they continue to face significant challenges, particularly security and privacy concerns when dealing with security-critical and sensitive data (design) \cite{mashnoor2025llmift, akyash2024selfhwdebug}\footnote{The fully closed nature of these commercial LLMs discourages their use on proprietary documents, design specifications, and data, as organizations are unwilling to risk the exposure of their confidential designs.}. Hence, the trend shows a shift from commercial to open-source LLMs for RTL code generation, driven by the need for greater control, customization, and privacy in specialized applications (e.g., fine-tuning LLM for RTL generation, specialized in AI accelerators) \cite{fu2023gpt4aigchip, akyash2024evolutionary}. Fig. \ref{fig:llm_rtl_tuning} shows a simplified top view of how fine-tuning and refinement processes adapt the open-source LLMs to become domain-specific, enabling it to effectively address hardware design automation challenges. 

Building on this general process of \textit{instruction-tuning}, as shown in Fig. \ref{fig:llm_rtl_tuning}, recent studies have developed several new open-source LLMs fine-tuned specifically for RTL code generation \cite{thakur2023verigen, liu2024rtlcodera, pei2024betterv, cui2024origen, gao2024autovcoder, liu2024craftrtl}. Table \ref{tab:comparison_motivation} summarizes the key features of each one of these RTL-oriented fine-tuned LLMs. As shown, with no multi-modal support, these methods rely solely on unstructured text, and are particularly prone to hallucinations due to a lack of structured context \cite{perozzi2024letgraph}, increasing the risk of generating instructions that inaccurately reflect the intent or functionality of the original code. This is while the retrieval-augmented generation (RAG) has proved substantial improvement by incorporating additional relevant information, which enriches the LLM's understanding and improves output alignment \cite{guu2020realm}. Often, this information takes the form of structured data, such as graphs, and recent research has investigated the ability of LLMs to understand and process these graph structures effectively \cite{luo2024graphinstruct, chai2023graphllm, he2024gretriever, fatemi2023talklikegraph, perozzi2024letgraph}. These studies have demonstrated that LLMs can effectively capture and interpret relationships and patterns within structured data which represent complex dependencies and hierarchies. Despite the growing interest in applying LLMs to graph-based data, their potential for analyzing RTL code remains almost unexplored. 

\begin{table*}[t]
\scriptsize
\setlength\tabcolsep{1.75pt}
\caption{Top View of Existing LLM-assisted Studies for RTL Code Generation.}
\label{tab:comparison_motivation}
\begin{tabular}{@{} l *{21}c @{}}
\toprule
Model & Key Novelty & Training Dataset, [Size] & Fine-tuned Model & Multi-modal & HW Efficiency & \\
\cmidrule(r){1-1}\cmidrule(r){2-2}\cmidrule(r){3-3}\cmidrule(r){4-4}\cmidrule(r){5-5}\cmidrule(r){6-6}
VeriGen \cite{thakur2023verigen} & \makecell{Fine-tuning on Dataset collected from \\ GitHub and Textbooks} & \makecell{Open-source, GitHub \\ and Textbooks, {[not listed]}} & CodeGen-16B & No & None \\
\cmidrule(r){1-1}\cmidrule(r){2-2}\cmidrule(r){3-3}\cmidrule(r){4-4}\cmidrule(r){5-5}\cmidrule(r){6-6}
RTLCoder \cite{liu2024rtlcodera} & \makecell{GPT-3.5-based Code-Instruction \\ Pair Synthesis} & \makecell{Open-source, \\ Synthesized, [27K]} & \makecell{Mistral-7B \\ DeepSeek-Coder-6.7b} & No & None \\
\cmidrule(r){1-1}\cmidrule(r){2-2}\cmidrule(r){3-3}\cmidrule(r){4-4}\cmidrule(r){5-5}\cmidrule(r){6-6}
BetterV \cite{pei2024betterv} &  \makecell{Applying Controllable Text \\ Generation w/ Discriminators for \\ Engineering Optimization} & \makecell{Closed-source, \\From internet, \\ {[not listed]}} & \makecell{CodeLlama-7B \\ DeepSeek-Coder-6.7b-Instruct \\ Code Qwen1.5-7B} & No & Area Improvement\\
\cmidrule(r){1-1}\cmidrule(r){2-2}\cmidrule(r){3-3}\cmidrule(r){4-4}\cmidrule(r){5-5}\cmidrule(r){6-6}
OriGen \cite{cui2024origen} & \makecell{Code-to-code Augmentation, \\ Self-reflection for Fixing} & Open-source, [222K] & DeepSeek-Coder-7B & No & \makecell{Iterative Functional \\ Correctness Check} \\
\cmidrule(r){1-1}\cmidrule(r){2-2}\cmidrule(r){3-3}\cmidrule(r){4-4}\cmidrule(r){5-5}\cmidrule(r){6-6}
AutoVCoder \cite{gao2024autovcoder} & \makecell{Domain-specific RAG with \\ Two-round LLM fine-tuning \\ for Constructive Prompting} & \makecell{Collected from Github, \\ {[not listed]}} & \makecell{Codellama-7B \\ DeepSeek-Coder-6.7B \\ CodeQwen1.5-7B} & \makecell{No \\ (Text and Retrieval$^{*}$)} & RAG-based Optimization \\
\cmidrule(r){1-1}\cmidrule(r){2-2}\cmidrule(r){3-3}\cmidrule(r){4-4}\cmidrule(r){5-5}\cmidrule(r){6-6}
CodeV \cite{zhao2024codev} & \makecell{Multi-Level Summarization \\ for Verilog Generation}& \makecell{Close-source \\ Github, [165K]} &\makecell{Codellama-7B \\ DeepSeek-Coder-6.7B \\ CodeQwen1.5-7B}& No & Code Generation Improvement \\ 
\cmidrule(r){1-1}\cmidrule(r){2-2}\cmidrule(r){3-3}\cmidrule(r){4-4}\cmidrule(r){5-5}\cmidrule(r){6-6}
CraftRTL \cite{liu2024craftrtl} & Correct-by-construction data & \makecell{Synthetic and GitHub, \\ {[80.1K]}} & \makecell{Codellama-7B \\ DeepSeek-Coder-6.7B \\ Starcoder2-15B} & No & Fine-tuning Correction \\
\cmidrule[.8pt](r){1-1}\cmidrule[.8pt](r){2-2}\cmidrule[.8pt](r){3-3}\cmidrule[.8pt](r){4-4}\cmidrule[.8pt](r){5-5}\cmidrule[.8pt](r){6-6}
\cmidrule(r){1-1}\cmidrule(r){2-2}\cmidrule(r){3-3}\cmidrule(r){4-4}\cmidrule(r){5-5}\cmidrule(r){6-6}
\textbf{\texttt{RTL++}} & \makecell{\textbf{Structural-based Optimization} \\ \textbf{(Graph Embedding for Instruction Tuning)}} & \textbf{Open-source, [200K]} & \textbf{Codellama-7B} & \makecell{\textbf{Yes} \\\textbf{(Graph and Text)}} & \makecell{\textbf{Structural Optimization,} \\ \textbf{Area and Delay Improvement}}\\
\midrule
\multicolumn{5}{l}{$^{*}$: This RAG is to identify the piece of data (RTL code). It has nothing to do with cross-modality understanding.}\\
\bottomrule
\vspace{-20pt}
\end{tabular}
\end{table*}

As LLMs struggle with hardware design due to their limited understanding of RTL's concurrent nature, representing code as graphs can address this by capturing the hierarchy, dependencies, and relationships between components. This structured encoding improves context awareness, enabling LLMs to produce outputs that are more accurate and aligned with the intended functionality of the design. To address this need, this paper introduces \texttt{RTL++}, the first multi-modal graph-augmented fine-tuned LLM designed for enhanced RTL code generation. In \texttt{RTL++}, a unique graph-based representation of designs will be incorporated as an supporting embedding during the instruction generation and fine-tuning phases, which improves both the model's functional and structural understanding. The main contributions of \texttt{RTL++} are as follows:

\begin{figure*}[b]
    \centering
    \vspace{-15pt}
    \includegraphics[width=0.9\textwidth]{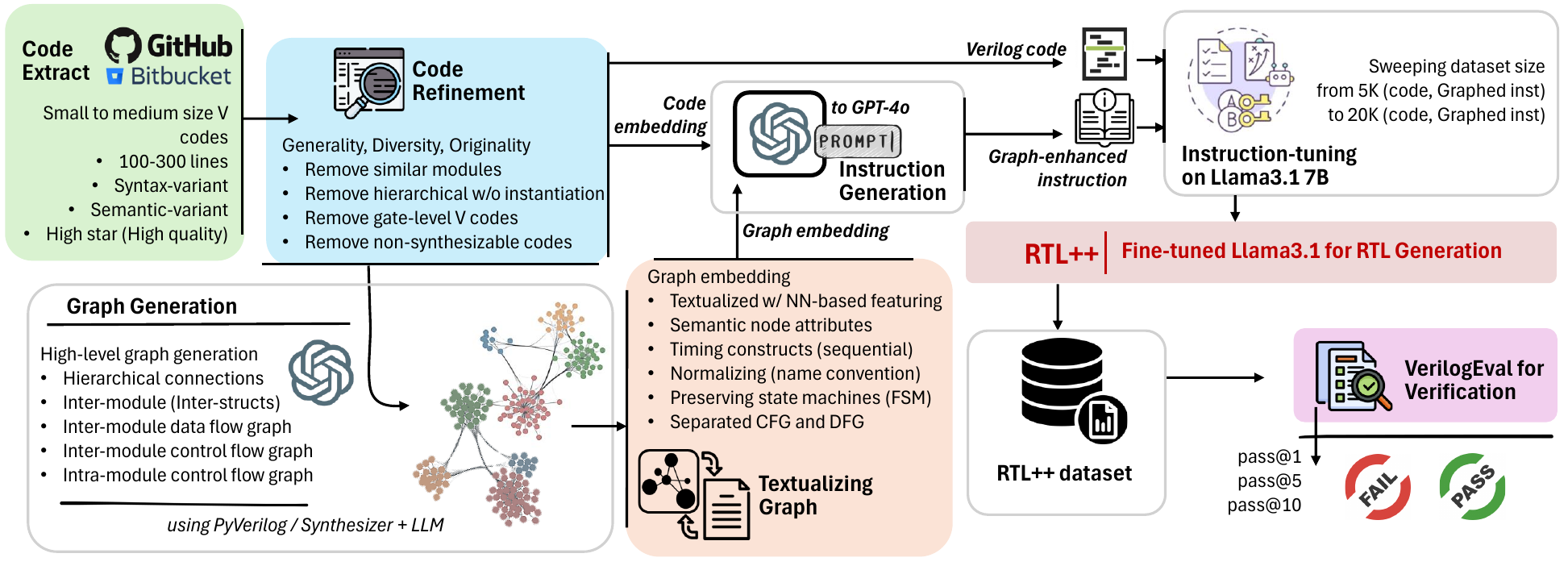}
    \caption{\texttt{RTL++} Overview: Text and Graph Embedding for LLM Fine-Tuning for RTL Code Generation.}
    \label{fig:proposed_overview}
\end{figure*}

\noindent (1) With an automated RTL-to-graph mechanism, which encodes each training dataset entry into \textit{control flow graphs} (CFG) and \textit{data flow graphs} (DFG), we introduce a new fine-tuning mechanism that relies on instructions generated based not only on the RTL code itself but also on its corresponding CFG and DFG\footnote{While RTL code (text) is for syntactical and semantical perspective, its CFG and DFG is for structural perspective of the circuit. This is conceptually a cross-modality fusion, in which the model integrate insights from both the RTL code (syntax and semantic) and its graph structures (structure).}. To the best of our knowledge, \texttt{RTL++} is the first LLM-assisted RTL code generator to enhance LLM efficiency by combining multiple data formats.

\noindent (2) In \texttt{RTL++}, a 100K training dataset has been curated using well-established open-source repositories such as GitHub, Bitbucket, and Opencores. The dataset entries are diverse, high-star rated, and contain critical keywords/structs\footnote{Structs include (but not limited to) \texttt{module}, \texttt{port}, \texttt{wire}, \texttt{reg}, procedural blocks (e.g., \texttt{always} and \texttt{initial}), control flow constructs (e.g., \texttt{if-else} and \texttt{case}), instantiation, FSM, parameters and constants, \texttt{generate} constructs, arrays and memories, etc.} in RTL generation. This diversity is crucial for LLM fine-tuning that can avoid overfitting to specific patterns. Also, to improve code quality, an LLM-assisted refinement (pruning) has been implemented to refine and optimize the collected code samples.

\noindent (3) We evaluated the latest foundational and advanced models, compared with our new model, highlighting that our multi-modal fine-tuning strategy sets a new benchmark in RTL coding. We also plan to make \texttt{RTL++} a fully open-source model with its 100K instruction tuning dataset to support collaboration in EDA and chip design community.

\section{Related Work}

\subsection{LLM for RTL Code Generation}

While recent advancements in hardware design automation have shown the effectiveness of adapting LLMs for specialized tasks like EDA automation and optimization, e.g., scripting \cite{liu2024chipnemo}, error interpretation \cite{chang2024data}, assistant chatbot for design flow \cite{wu2024chateda}, etc., numerous efforts have been made to fine-tune and pre-train models for RTL (Verilog) code generation: 

\noindent \underline{(i) \textbf{\textit{VeriGen}}} \cite{thakur2023verigen} is an early attempt that compiled Verilog files from GitHub and textbooks for training dataset. Despite assembling a substantial dataset, the lack of proper pre-processing and organization led to inconsistencies which causes the fine-tuned CodeGen model to often generate non-functional Verilog code with syntax errors. 

\noindent \underline{(ii) \textbf{\textit{RTLCoder}}} \cite{liu2024rtlcodera} is another early endeavor that has leveraged GPT-3.5 for synthesizing code-instruction pairs by extracting RTL-specific keywords (to overcome dataset scarsity and code generation quality).  However, its dependence on GPT-3.5’s embedded knowledge limited code diversity. This limitation prompted subsequent efforts to enrich data diversity through augmentation techniques. 

\noindent \underline{(iii) \textbf{\textit{OriGen}}}  \cite{cui2024origen} advanced RTL code generation by introducing code-to-code augmentation and a self-reflection mechanism. The former diversifies the dataset with semantically equivalent but syntactically varied Verilog code, while the latter uses compiler feedback to iteratively correct errors, addressing VeriGen's inconsistencies and improving code quality.

\noindent \underline{(iv) \textbf{\textit{BetterV}}} \cite{pei2024betterv}, building upon OriGen's idea of augmenting data and integrating feedback, extends the capabilities of Verilog code generation by introducing a controlled text generation framework tailored specifically for Verilog, drawing parallels with C programs to help LLMs better comprehend Verilog semantics. It employs generative discriminators to optimize the Verilog for Power, Performance, and Area (PPA) while also incorporating data augmentation techniques to address data scarcity issues. 

\noindent \underline{(v) \textbf{\textit{AutoVCoder}}} \cite{gao2024autovcoder} focuses on addressing the limitation of diversity and domain-specific accuracy in RTL code generation using a two-round fine-tuning process to boost LLM performance in Verilog code generation. AutoVCoder also incorporated a domain-specific RAG module to constructively enhance prompts, which improved the syntactic and functional correctness of the generated code. 

\noindent \underline{(vi) \textbf{\textit{CodeV}}} \cite{zhao2024codev} leverages LLMs for code summarization rather than generation, shifting the focus from producing Verilog code from natural language to generating detailed descriptions from Verilog code. By processing 165K Verilog modules from GitHub and fine-tuning with multi-level summarization, CodeV enhances training datasets with rich description-code pairs, ensuring both syntactic accuracy and semantic depth for high-quality Verilog representations.

\noindent \underline{(vii) \textbf{\textit{CraftRTL}}} \cite{liu2024craftrtl} introduces an approach that includes constructing correct-by-construction data, such as Karnaugh Maps, state-transition diagrams, and waveforms, which improve the ability of LLMs to interpret additional information for LLM fine-tuning. In addition, CraftRTL employs an automated framework that uses LLMs to generate error reports at various training checkpoints.

MAGE \cite{zhao2024mage} enhances RTL generation using a multi-agent system with high-temperature sampling and Verilog-state checkpointing, but for evaluation, we focus on approaches that improve model performance through fine-tuning and data collection. While recent advancements in RTL code generation have shown promising improvements, they lack (i) \textbf{capturing the hierarchical structure of designs},  (ii) \textbf{comprehending data and control flow}, and (iii) \textbf{addressing the intrinsic concurrency} of hardware designs. Prior NLP-based studies \cite{cai2024codegraph, luo2024reasoninggraphs, perozzi2024letgraph} have demonstrated the power of integrating structured knowledge to enhance reasoning and boost interpretability in LLMs. Inspired by these works, our proposed \texttt{RTL++} incorporates graph-based knowledge of RTL design to bridge the gap between design abstraction and design phase.

\subsection{Graph Prompt Learning and Engineering}

Since the advent of LLMs, researchers have been exploring ways to embed graph data into the input of LLMs (as an embedding to use as in-context learning) to enable reasoning over graph-structured information \cite{luo2024graph, chai2023graphllm, he2024gretriever,luo2024reasoninggraphs, perozzi2024letgraph, fatemi2023talklikegraph, ren2022graph}. 

Fatemi et al. \cite{fatemi2023talklikegraph} evaluated the encoding graph-structured data as text for LLMs. A key observation of this study is that LLM performance on graph reasoning tasks is highly sensitive to the chosen encoding method, the type of task, and the structure of the graph, hence emphasizing on selecting appropriate graph encoding techniques is paramount. Perozzi et al. \cite{perozzi2024letgraph} proposed a novel approach that leverages graph neural networks (GNNs) to encode data into embeddings instead of textualizing graphs. They introduced GraphToken, a parameter-efficient method explicitly designed to encode structured graph data for LLMs. GraphToken learns an encoding function that augments prompts with explicit structured information. Unlike GraphToken, GraphLLM \cite{chai2023graphllm} adopts an end-to-end approach, integrating graph learning models with LLMs. It employs a graph transformer to process graph structures directly, enhancing both accuracy and efficiency.

Inspired by these approaches, we also leverage textualized graph representations to integrate graph data into LLMs. Specifically, we convert graph-structured data from RTL codes, such as CFG and DFG, into textual formats, which are then used alongside RTL codes to generate meaningful instructions. By providing both the graph and corresponding code, we enable LLMs to effectively reason about and generate complex hardware design instructions.

\section{Proposed Model: \texttt{RTL++}}

Fig. \ref{fig:proposed_overview} presents a top-down view of our proposed model, \texttt{RTL++}, which is structured into five key steps: (1) RTL code collection, (2) RTL code refinement, (3) RTL CFG/DFG generation, (4) instruction generation (based on RTL code and CFG/DFG), (5) graph-enhanced instruction-tuning. These steps, each detailed below, collectively prepare the fine-tuned LLM for high-quality RTL code generation.

\subsection{Collecting RTL Dataset from Repositories}
\label{subsec:collection}

To gather a high-quality RTL code dataset, i.e., from GitHub, Bitbucket, and Opencores, we targeted a list of most common keywords relevant to RTL design (see Table \ref{tab:keyword_targeted}. For each keyword, we generated 10 related sub-keywords to capture a wider range of code (maximizing diversity). This keyword expansion allowed us to cover more specific design cases, including real common use cases in hardware design. We filtered out data with less than 100 lines and more than 300 lines to maintain consistency and focus on small-sized designs, which balance complexity and make them ideal for realistic use cases and model training.

To ensure that we selected the high quality and reliable RTL codes, we ranked GitHub repositories by their star count, assuming that more popular repositories—indicated by a higher number of stars—likely contain well-maintained, reliable code. From these top-ranked repositories, we extracted RTL code files with high star counts, as this metric often correlates with quality and community validation.

\begin{table}[t]
\scriptsize
\centering
\caption{Main Targeted Keyword in Dataset Collection in \texttt{RTL++}}
\label{tab:keyword_targeted}
\setlength\tabcolsep{2pt} 
\begin{tabular}{@{} p{50pt} p{189pt}@{}}
\toprule  
\textbf{Category} & \textbf{Struct-based Keywords} \\
\cmidrule[0.85pt](r){1-1}\cmidrule[0.85pt](r){2-2}
Structural ~~~~~Constructs & \texttt{module}, \texttt{endmodule}, \texttt{input}, \texttt{output}, \texttt{inout}, \texttt{wire}, \texttt{reg}, \texttt{assign}, \texttt{generate}, \texttt{endgenerate}, \texttt{parameter}, \texttt{localparam}, \texttt{always @(*)}, \texttt{begin}, \texttt{end} \\
\cmidrule(r){1-1}\cmidrule(r){2-2}
Sequential ~~~~~Logic & \texttt{always}, \texttt{always @(posedge clk)}, \texttt{always @(negedge clk)}, \texttt{posedge}, \texttt{negedge}, \texttt{if}, \texttt{else}, \texttt{case}, \texttt{default}\\ 
\cmidrule(r){1-1}\cmidrule(r){2-2}
Combinational Logic & \texttt{assign}, \texttt{case}, \texttt{casex}, \texttt{casez}, \texttt{and}, \texttt{or}, \texttt{not}, \texttt{nand}, \texttt{nor}, \texttt{xor}, \texttt{xnor}, \texttt{mux}, \texttt{demux}, \texttt{generate}, \texttt{genvar} \\ 
\cmidrule(r){1-1}\cmidrule(r){2-2}
Memory ~~~~~~~~~and Storage & \texttt{always @(posedge clk or negedge reset)}, Flip-flop constructs (e.g., \texttt{if(enable)}), \texttt{RAM}, \texttt{ROM}, \texttt{initial} \\ 
\cmidrule(r){1-1}\cmidrule(r){2-2}
Data Path & \texttt{add}, \texttt{sub}, \texttt{mul}, \texttt{div}, \texttt{\textless\textless}, \texttt{\textgreater\textgreater}, \texttt{+}, \texttt{-}, \texttt{*}, \texttt{/}, \texttt{\&}, \texttt{|}, \texttt{\^{}}, \texttt{\~{}} \\
\cmidrule(r){1-1}\cmidrule(r){2-2}
State ~~~~~~~~~~~~~~~~Machines & \texttt{always @(posedge clk)}, \texttt{case}, \texttt{endcase}, \texttt{default}, \texttt{parameter}, \texttt{localparam}, \texttt{idle}, \texttt{current}, \texttt{next}  \\ 
\cmidrule(r){1-1}\cmidrule(r){2-2}
Hierarchy & \texttt{module instantiation}, Dot (\texttt{.port\_name(xyz)}) \\ 
\bottomrule
\toprule  
\textbf{Category} & \textbf{Context-based Keywords} \\
\cmidrule[0.85pt](r){1-1}\cmidrule[0.85pt](r){2-2}
Arith \& Logic & \texttt{adder}, \texttt{subtractor}, \texttt{multiplier}, \texttt{divider}, \texttt{alu} \\
\hline
Sequential Logic & \texttt{dff}, \texttt{register}, \texttt{shift\_register}, \texttt{counter} \\
\hline
Memory & \texttt{register\_file}, \texttt{register\_bank}, \texttt{fifo}, \texttt{cache} \\
\hline
Control Logic &  \texttt{encoder}, \texttt{decoder}, \texttt{arbiter}, \texttt{bus\_controller} \\
\hline
Data Transfer & \texttt{uart}, \texttt{spi}, \texttt{i2c}, \texttt{ethernet}, \texttt{input\_buffer}, \texttt{output\_buffer}, \texttt{parity}, \texttt{hamming}, \texttt{crc} \\
\hline
Signal Proc. & \texttt{filter}, \texttt{fft}, \texttt{dft}, \texttt{mac}, \texttt{cordic} \\
\hline
Interconnect & \texttt{axi}, \texttt{wishbone}, \texttt{apb}, \texttt{crossbar}, \texttt{bus\_switch}, \texttt{bridge} \\
\hline
Clock & \texttt{pll}, \texttt{clock\_divider}, \texttt{prescaler}, \texttt{timer}, \texttt{stopwatch} \\
\bottomrule
\end{tabular}
\vspace{-15pt}
\end{table}

In addition to ranking by popularity, we filtered out testbenches and netlists, focusing exclusively on RTL (behavioral) design files. We also ensured that the collected RTL codes contained all \texttt{module}s with their needed decalaration (for hierarchical design). By considering these extra steps for RTL code collection, we create a clean, relevant dataset of RTL design code suitable for training purposes.

\subsection{RTL Code Refinement Using GPT}

While state-of-the-art studies focuses on using either machine-generated (LLM-based) or human-created RTL code for training datasets, in \texttt{RTL++}, we employ a hybrid approach, where we collect RTL codes ($\S$\ref{subsec:collection}) and refine them using the machine (here GPT-4o). The refinement process involved a structured prompt that guided GPT through several key steps to standardize and enhance each RTL module (making the RTL code consistent, syntactically correct, self-contained, and aligned with the requirements for effective model training):

\noindent (i) \textit{\underline{Dependency Removal:}} For modules dependent on external files or submodules instantiated within the main module\footnote{While we exclude code dependencies on external sources (at $\S\ref{subsec:collection}$), GPT-based auto-completion is used to fix incomplete code fragments.}, GPT embedded the logic of these dependencies directly within the code. External module instantiations were replaced by their corresponding internal logic, creating self-contained modules that no longer required external dependencies. 

\noindent (ii) \textit{\underline{Variable Definition and Initialization:}} In large RTL projects, libraries or headers often define global variables. However, when assembling sub-modules from these hierarchical projects, these reference files may not always be included in the training dataset. So, GPT is called to set value to these undefined variables and functions (infers typical use cases or context-based values), creating self-contained, standalone modules for enhanced training utility.

\noindent (ii) \textit{\underline{Syntax Check/Correction and Synthesizability:}} GPT is invoked to check\footnote{Each sample has \texttt{module} and \texttt{endmodule} with a procedural block (e.g., \texttt{always@...}, \texttt{assign}, \texttt{.instance(port)} , while inputs (\texttt{in/inout} ports) are connected to outputs (\texttt{out/inout} ports).}/correct syntactical structures to adhere to standard RTL (Verilog) syntax, addressing common elements such as operator usage, and supported constructs. Additionally, to validate the correctness of the collected and refined RTL codes, a basic synthesis flow was performed using Yosys \cite{wolf2013yosys}\footnote{Codes with synthesis error are excluded from the further steps and training.}.

\subsection{High-level Graph Generation from RTL Codes}

There are various methods to encode text into graphs for LLMs, such as using GNNs \cite{perozzi2024letgraph}, and graph convolutional networks (GCNs) \cite{ren2022graph}, and directly integrating them with LLMs. Some representations are particularly well-suited for LLMs as they balance structural accuracy with textual clarity \cite{fatemi2023talklikegraph}. In \texttt{RTL++}, we employ textual descriptions of graphs because they offer better interpretability for LLMs, making it easier for the models to understand hierarchies. By potentially leveraging these descriptions as RAGs, we enable LLMs to tackle more complex reasoning tasks in auto-debugging, optimization, and verification.

We follow these steps to generate textual graphs in \texttt{RTL++}: 

\noindent (1) Flattening (embedding all modules into the main graph) that is for designs with instantiated modules (hierarchical). 

\noindent (2) DFG generation (module-level to I/O-level) that is for data movement form inputs to outputs.

\noindent (3) CFG generation (module-level to I/O-level) that is for control signals form inputs to outputs.

\noindent (4) Adding node attributes that includes type (gate, module, input, output, etc.) and function (arith, storage, logic, etc.).

\noindent (5) Adding temporal behavior of nodes that determines the sequence of operation (based on sequencing\footnote{BMC engines can be used for iterations to collect sequencing \cite{clarke2001bounded}.}). It also includes feedback loops in sequential circuits (e.g., FSMs).

Using these steps in \texttt{RTL++}, we design all graphs at a high level, keeping details minimal to maintain a balance between abstraction and usability. Our primary focus is on defining graphs at the module level, avoiding unnecessary details that could overwhelm the model. This approach aligns closely with how hardware engineers conceptualize RTL code.

\vspace{-4pt}

\subsection{Graph-enhanced Instruction Generation}

To create code-instruction pairs for fine-tuning in \texttt{RTL++}, relying on in-context learning \cite{dong2022survey}, the prompting includes both code snippets and their associated (textualized) graph-based representations. By integrating detailed information from both the code and its graphical representations, the LLM generates instructions that are more accurate, informative, and aligned with the actual functionality of the hardware module. The instructions encapsulate complex control flows and data interactions (from CFG and DFG) in a clear and concise manner. By using CFG and DFG as additional embeddings, the LLM enhances its understanding of critical component interactions and the intended purpose of various modules. This results in RTL code that is more detailed, precise, and less likely to include inaccuracies or hallucinations.

\vspace{-4pt}

\subsection{Fine-Tuning over Code-Instruction Pairs}

At the end, once the pairs of RTL codes and (graph-enghanced) instructions are ready as our dataset, we finetune the base model (i.e., CodeLlama \cite{rozière2024codellama}) on this dataset\footnote{While DeepSeek \cite{guo2024deepseek} could obtain superior outcomes as the base model, we opted for CodeLlama \cite{rozière2024codellama} to evaluate the genuine impact of incorporating graph structures during fine-tuning.}.

\begin{table}[t]
\scriptsize
\setlength\tabcolsep{3pt}
\caption{\texttt{RTL++} vs. base CodeLlama-7B-instruct and GPT-4.}
\label{tab:basic_perf}
\begin{tabular}{@{} l *{21}c @{}}
\toprule
\multirow{3}{*}{Evaluated Model} & \multirow{3}{*}{no. of params} & \multirow{3}{*}{Open-Source?} & \multicolumn{3}{c}{VerilogEval (pass@k) \cite{Liu2023verilogeval}}\\
\cmidrule(r){4-6}
& & & \multicolumn{3}{c}{Only HumanEval$^*$} \\
\cmidrule(r){4-6}
& & & k = 1 & k = 5 & k = 10 \\
\cmidrule(r){1-1}\cmidrule(r){2-2}\cmidrule(r){3-3}\cmidrule(r){4-4}\cmidrule(r){5-5}\cmidrule(r){6-6}
GPT-4 & N/A & \xmark & 43.5 & 55.8 & N/A \\
\cmidrule(r){1-1}\cmidrule(r){2-2}\cmidrule(r){3-3}\cmidrule(r){4-4}\cmidrule(r){5-5}\cmidrule(r){6-6}
CodeLlama-based & 7B & \cmark & 18.2 & 22.7 & 24.3\\ 
\cmidrule(r){1-1}\cmidrule(r){2-2}\cmidrule(r){3-3}\cmidrule(r){4-4}\cmidrule(r){5-5}\cmidrule(r){6-6}
\texttt{RTL++} @ 5K Trained  & \multirow{7}{*}{7B} & \multirow{7}{*}{\cmark}& 23.7 & 28.2 & 30.7 \\
\texttt{RTL++} @ 10K Trained &&&  26.2 & 30.1 & 33.3 \\ 
\texttt{RTL++} @ 15K Trained &&&  28.2 & 32.6 & 34.6 \\ 
\texttt{RTL++} @ 20K Trained &&& 29.4 & 34.6 & 37.1 \\
\texttt{RTL++} @ 50K Trained &&& 41.3 & 47.1 & 50.5 \\
\textbf{\texttt{RTL++} @ 100K Trained} &&& \textbf{\underline{54.3}} & \textbf{\underline{60.8}} & \textbf{\underline{65.2}} \\
\textbf{\texttt{RTL++} @ 200K Trained} &&& \textbf{\underline{59.9}} & \textbf{\underline{68.8}} & \textbf{\underline{72.1}}\\
\bottomrule
\multicolumn{5}{l}{$^*$: HumanEval ensures RTL evaluation aligns with real-world data \cite{pinckney2024revisiting}.}
\end{tabular}
\vspace{-15pt}
\end{table}

\section{Experiments}

To evaluate the performance of \texttt{RTL++}, we fine-tune the CodeLlama-7B-Instruct as the targeted generative LLM. All experiments are conducted for 1 epoch (to avoid over-fitting, as we observed over-fitting when using more epochs) using PyTorch on NVIDIA L4 with the learning rate at 2e-4. Additionally, for RTL code refinement, graph refinement, and instruction generation, GPT-4 has been engaged, costing approximately \$84 per 1000 samples.

To assess the \texttt{RTL++} performance, we utilized two benchmarking frameworks: VerilogEval\footnote{Built upon the revisited VerilogEval \cite{pinckney2024revisiting}} \cite{Liu2023verilogeval} and RTLLM \cite{lu2023rtllm}. Both benchmarks employ the widely recognized pass@k\footnote{Pass@k is the percentage of problems solved within k attempts.} metric to evaluate the correctness of the generated code.

To fine-tune the model for RTL code generation, we leverage the LoRA (Low-Rank Adaptation) \cite{hu2021lora} technique (enhancing RTL-oriented capabilities while maintaining performance). The optimization process employs the AdamW optimizer, configured with $\beta_1 = 0.9$ and $\beta_2 = 0.99$, and a cosine decay schedule for the learning rate. A warm-up phase is included with a ratio of 0.03, and training batch size is 2.

We compared \texttt{RTL++} against several existing models, including Verigen \cite{thakur2023verigen}, RTLCoder \cite{liu2024rtlcodera}, BetterV \cite{pei2024betterv}, Origen \cite{cui2024origen}, AutoVCoder \cite{gao2024autovcoder} and CraftRTL \cite{liu2024craftrtl}. Additionally, CodeLlama-7B-instruct was used as a baseline to assess the improvements made by \texttt{RTL++}, while comparison with GPT-4 has been also explored (to show \texttt{RTL++} efficiency).

\vspace{-5pt}

\subsection{Comparison with the Base Models}

Table \ref{tab:basic_perf} shows the performance comparison between \texttt{RTL++}, the base CodeLlama-7B-instruct model, and GPT-4. The results indicate that as the training dataset for \texttt{RTL++} grows, the quality of the generated code consistently improves. Notably, \texttt{RTL++} outperforms GPT-4 when trained on a 100K dataset. Note that for the VerilogEval benchmark, we prioritize HumanEval as it more accurately reflects real-world data \cite{pinckney2024revisiting}.

\vspace{-3pt}

\subsection{Impact of data size}

\begin{figure}
    \centering
    \includegraphics[width=\columnwidth]{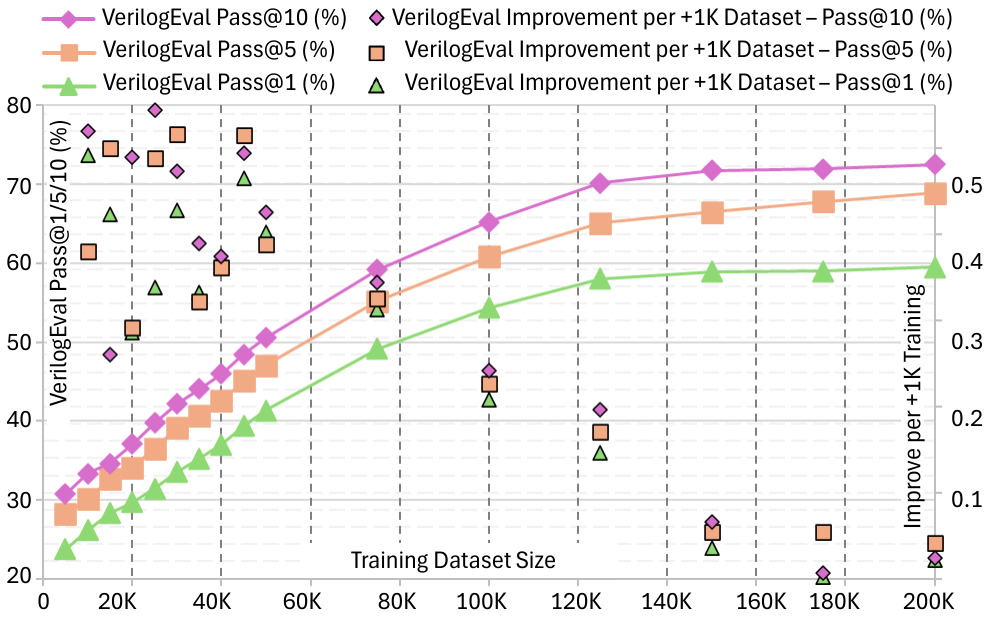}
    \caption{Impact of dataset size on model accuracy.}
    \label{fig:data_vs_accuracy}
    \vspace{-5pt}
\end{figure}

The impact of dataset size on the accuracy of our model is illustrated in Figure \ref{fig:data_vs_accuracy}. We trained the model using datasets collected up to 200K samples. This suggests that if we gather more data, we can achieve higher accuracy levels. The initial collected data (5K, 10K, and 15K) showed a gradual improvement in model accuracy, but with larger datasets, there is a clear trend of significant performance gains. The combination of high-quality data, augmenting instruction generation by graphs, and effective model training demonstrate the potential to achieve even higher accuracy levels with larger datasets.

\vspace{-3pt}

\subsection{Comparison with the State-of-the-art Models}

Table \ref{tab:whole_comparison} presents a comprehensive comparison between the performance of \texttt{RTL++} and other state-of-the-art models. While many competing models leverage both CodeLlama \cite{rozière2024codellama} and DeepSeek \cite{guo2024deepseek}, all reported results are based on fine-tuned versions of CodeLlama-7B-Instruct to ensure a consistent and fair evaluation. The results demonstrate that, with an expanded training dataset, \texttt{RTL++} outperforms these models, highlighting the critical role of multi-modal embedding, particularly augmenting the embedding using CFGs and DFGs, in improving LLM-assisted RTL code generation.

\begin{table}[t]
\scriptsize
\setlength\tabcolsep{3pt}
\caption{Comparsion of \texttt{RTL++} and state-of-the-art RTL Generators.}
\label{tab:whole_comparison}
\begin{tabular}{@{} l *{21}c @{}}
\toprule
\multirow{3}{*}{Evaluated Model$^*$} & \multirow{3}{*}{\#ps} & \multirow{3}{*}{OSS?} & \multirow{3}{*}{Tr Size} & \multicolumn{3}{c}{VerilogEval \cite{Liu2023verilogeval}} & \multicolumn{2}{c}{RTLLM 1.1 (@5)} \\
\cmidrule(r){5-7} \cmidrule(r){8-9}
& & & & \multicolumn{3}{c}{HumanEval (\%)} & \multirow{2}{*}{Syn.} & \multirow{2}{*}{Func.} \\
\cmidrule(r){5-7} 
& & & & @1 & @5 & @10 \\
\cmidrule(r){1-1}\cmidrule(r){2-2}\cmidrule(r){3-3}\cmidrule(r){4-4}\cmidrule(r){5-5}\cmidrule(r){6-6}\cmidrule(r){7-7}\cmidrule(r){8-8}\cmidrule(r){9-9}
GPT-4 & N/A &\xmark & - &  43.5 & 55.8 & N/A & 89.7 & 37.9\\
\cmidrule(r){1-1}\cmidrule(r){2-2}\cmidrule(r){3-3}\cmidrule(r){4-4}\cmidrule(r){5-5}\cmidrule(r){6-6}\cmidrule(r){7-7}\cmidrule(r){8-8}\cmidrule(r){9-9}
CodeLlama-7B-I & 7B & \cmark & N/A & 18.2 & 22.7 & 24.3 & 62.1 & 10.3 \\  
\cmidrule(r){1-1}\cmidrule(r){2-2}\cmidrule(r){3-3}\cmidrule(r){4-4}\cmidrule(r){5-5}\cmidrule(r){6-6}\cmidrule(r){7-7}\cmidrule(r){8-8}\cmidrule(r){9-9}
VeriGen \cite{thakur2023verigen} & 15B & \cmark & - &30.3 &43.9 &49.6& 86.2 & 24.1 \\ 
RTLCoder-DS \cite{liu2024rtlcodera} & 6.7B & \cmark & 27K & 41.6 & 50.1 & 53.4  &93.1 & 48.3 \\
BetterV-CQ \cite{pei2024betterv}  & 7B & \xmark & - & 46.1 & 53.7 & 58.2 & N/A & N/A \\
OriGen \cite{cui2024origen} & 7B & \xmark & 222K & 54.4 &  60.1 & 64.2 & N/A & 65.5\\
AutoVCoder-CQ \cite{gao2024autovcoder} & 7B & \xmark & - & 48.5 & 55.9 & N/A & 100 & 51.7 \\
CodeV \cite{zhao2024codev} & 7B & \cmark & 165K & 45.2 & 59.5 & 63.8 & 93.1 & \textbf{\underline{62.1}}\\
CraftRTL \cite{liu2024craftrtl} & 7B & \xmark & 80.1K  & \textbf{\underline{63.1}} & 67.8 &69.7 & 93.9 & 52.9\\
\cmidrule(r){1-1}\cmidrule(r){2-2}\cmidrule(r){3-3}\cmidrule(r){4-4}\cmidrule(r){5-5}\cmidrule(r){6-6}\cmidrule(r){7-7}\cmidrule(r){8-8}\cmidrule(r){9-9}
\textbf{\texttt{RTL++}@50K}  & \multirow{3}{*}{7B} & \multirow{3}{*}{\cmark} & 50K &  41.3 & 47.1 & 50.5 & 82.7 & 41.3\\
\textbf{\texttt{RTL++}@100K} &&& 100K & 54.3 & 60.8 & 65.2 & 86.2 & 44.8\\ 
\textbf{\texttt{RTL++}@200K} &&& 200K & 59.9 & \textbf{\underline{68.8}} & \textbf{\underline{72.1}} & \textbf{\underline{93.9}} & 51.7\\
\bottomrule
\multicolumn{8}{l}{$^*$: Our focus is on works that involve both data collection and fine-tuning.
}
\end{tabular}
\vspace{-15pt}
\end{table}

\begin{table}[b]
\scriptsize
\setlength\tabcolsep{2.5pt}
\vspace{-10pt}
\caption{Ablation Study on Textualized Graph Rep. (TGR) @ 5K Datasets.}
\label{tab:ablation_no_graph}
\begin{tabular}{@{} l *{21}c @{}}
\cmidrule[0.85pt](r){1-5} \cmidrule[0.85pt](r){6-10}
\multirow{3}{*}{Model} & \multirow{3}{*}{Temp.} & \multicolumn{3}{c}{VerilogEval \cite{Liu2023verilogeval}} & \multirow{3}{*}{Model} & \multirow{3}{*}{Temp.} & \multicolumn{3}{c}{VerilogEval \cite{Liu2023verilogeval}} \\
\cmidrule(r){3-5} \cmidrule(r){8-10}
& & @1 & @5 & @10 & & & @1 & @5 & @10 \\
\cmidrule(r){1-1}\cmidrule(r){2-2}\cmidrule(r){3-5}\cmidrule(r){6-6}\cmidrule(r){7-7}\cmidrule(r){8-10}
\texttt{RTL++} w/o TGR & 0.6 & 22.4 & 23.7 & 25.6 & \texttt{RTL++} w/ TGR & 0.6 & 23.7 & 27.5 & 29.4\\
\cmidrule(r){1-1}\cmidrule(r){2-2}\cmidrule(r){3-5}\cmidrule(r){6-6}\cmidrule(r){7-7}\cmidrule(r){8-10}
\texttt{RTL++} w/o TGR & 0.7 & 22.4 & 24.3 & 25.6 & \texttt{RTL++} w/ TGR & 0.7 &  23.7 & 28.2 & 30.7 \\

\cmidrule[0.85pt](r){1-5} \cmidrule[0.85pt](r){6-10}
\end{tabular}
\end{table}

\vspace{-3pt}

\subsection{Impact of Graph Utilization on Instruction Generation}

To evaluate the impact of graph-based augmentation on instruction generation quality in \texttt{RTL++}, Table \ref{tab:ablation_no_graph} compares models trained with and without textualized graph representations. Training the model using instructions augmented with textualized graph representations results in a noticeable accuracy improvement. Table \ref{tab:ablation_no_graph} provides a representative ablation study based on a 5K dataset, showing up to a 5\% increase in pass@10. When scaled to a 100K dataset, the improvement grows significantly, reaching 18\%, underscoring the value of graph augmentation for fine-tuning.

\subsection{A Simple Case Study: An ALU in \texttt{RTL++}}

\begin{figure}[t]
    \centering
    \includegraphics[width=0.9\columnwidth]{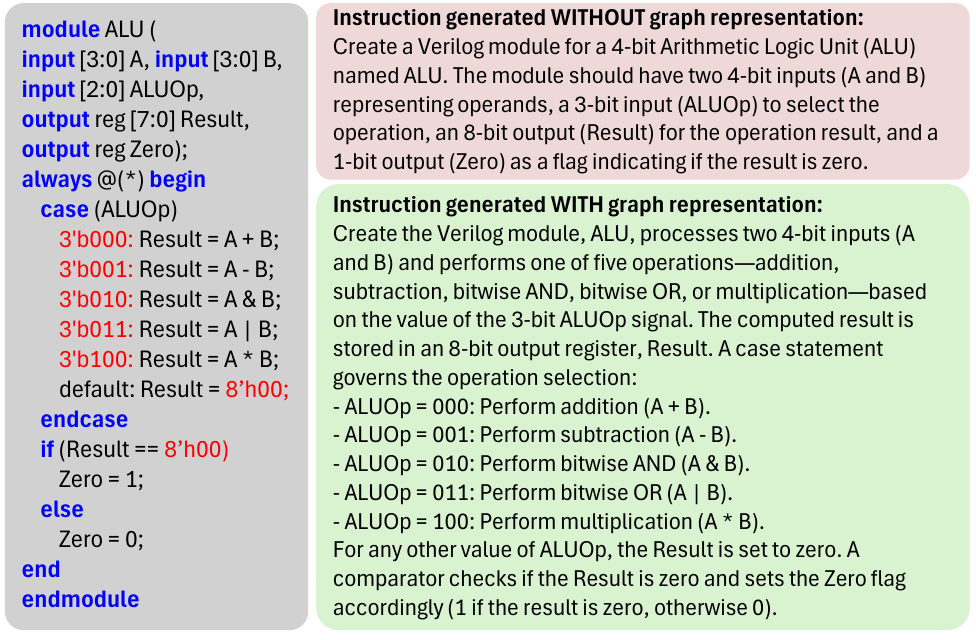}
    \caption{Instructions w/ and w/o CFG and DFG and corresponding RTL code.}
    \label{fig:graph_based}
    \vspace{-10pt}
\end{figure}

To gain deeper insights into the impact of graph-based instruction generation, we performed an ablation study focusing on a RTL module for an arithmetic logic unit (ALU). The ALU takes two 4-bit inputs ($A$ and $B$), performs a variety of arithmetic and logical operations based on a 3-bit control signal $(ALUOp)$, and outputs an 8-bit result (Result) along with a zero flag (Zero). As in Fig. \ref{fig:graph_based}, For this study, we generated instructions using two different configurations: one that included graph-based representations, as additional input, and one that used only the RTL code with no graph. As shown, the instruction with the graph representation provides a more technically detailed and complete description compared to the one without it. This version systematically defines the functionality of the 4-bit ALU by explicitly mapping ALUOp values to specific operations (addition, subtraction, bitwise AND, OR, and multiplication) through a case statement, while also handling invalid ALUOp cases by setting the Result to zero. It also clarifies the generation of the Zero flag, which indicates whether the computed Result is zero, ensuring robustness in implementation. In contrast, the instruction generated without the graph representation lacks this detailed description of control signal dependencies and default behaviors which leads to ambiguity. The inclusion of graph representation enhances clarity, depicting control flow, data dependencies, and interactions more precisely, which is crucial for accurate RTL code implementation.

In this specific example, the instruction generated with the both DFG and CFG representation closely aligns with the code, which can effectively capture the control flow and structural dependencies of the RTL module. By explicitly depicting the relationships between control signals and data, it provides a more structured understanding, which leads to better training outcomes for LLMs. 
\vspace{-2.5pt}
\section{Conclusion}

This paper introduces \texttt{RTL++}, an fully open-source model leveraging LLMs for efficient RTL code generation. \texttt{RTL++} is a first-of-its-kind that integrates both textual (RTL code) and graph-based representations (CFG and DFG in textualized formats) to generate high-quality instruction-code pairs for LLM fine-tuning for RTL generation purposes. By using multi-modal fine-tuning approach, \texttt{RTL++} achieves remarkable performance, surpassing state-of-the-art competitive models. The experimental results show that its success rate exceeds 70\% in VerilogEval and 90\% in RTLLM benchmark, all while operating on a fine-tuned version of CodeLlama-7B-Instruct.

\bibliographystyle{IEEEtran}
\bibliography{refs}

\end{document}